\documentclass[apjl]{emulateapj}

\def\etacar{$\eta$~Car }
\def\etacarp{$\eta$~Car}

\def\agl{1AGL~J1043-5931 }
\def\aglp{1AGL~J1043-5931}

\def\agll{AGL~J1046-5832 }

\def\rv{ }
\def\mt{ }
\def\mtav{  }
\def\mtavv{ }
\def\mtt{  }

\slugcomment{\bf ApJLetters, in press, 16 Mar. 2009}

\shorttitle{Detection of gamma-ray emission from the Eta Carinae
region} \shortauthors{Tavani et al.}

\begin{document}


\title{\bf Detection of Gamma-Ray Emission
    from the Eta-Carinae Region}


\bigskip

\author{
M.~Tavani\altaffilmark{1,2}, S. Sabatini\altaffilmark{2},
E.~Pian\altaffilmark{16}, A.~Bulgarelli\altaffilmark{5},
P.~Caraveo\altaffilmark{3}, R.F.~Viotti\altaffilmark{1},
M.F.~Corcoran\altaffilmark{17},
 A.~Giuliani\altaffilmark{3},
 C.~Pittori\altaffilmark{14},
 F.~Verrecchia\altaffilmark{14},
S.~Vercellone\altaffilmark{3}, S.~Mereghetti\altaffilmark{3},
 A.~Argan\altaffilmark{1},
G.~Barbiellini\altaffilmark{6}, F.~Boffelli\altaffilmark{7},
 P.~W.~Cattaneo\altaffilmark{7},
A.~W.~Chen\altaffilmark{3,4}, V.~Cocco\altaffilmark{1},
F.~D'Ammando\altaffilmark{1,2}, E.~Costa\altaffilmark{1}, G.~De
Paris\altaffilmark{1}, E.~Del Monte\altaffilmark{1},
G.~Di~Cocco\altaffilmark{5}, I.~Donnarumma\altaffilmark{1},
Y.~Evangelista\altaffilmark{1}, A.~Ferrari\altaffilmark{4,18}
M.~Feroci\altaffilmark{1}, M.~Fiorini\altaffilmark{3},
T.~Froysland\altaffilmark{2,4}, 
F.~Fuschino\altaffilmark{5}, M.~Galli\altaffilmark{8},
F.~Gianotti\altaffilmark{5}, C.~Labanti\altaffilmark{5},
I.~Lapshov\altaffilmark{1}, F.~Lazzarotto\altaffilmark{1},
P.~Lipari\altaffilmark{9}, F.~Longo\altaffilmark{6},
M.~Marisaldi\altaffilmark{5}, M.~Mastropietro\altaffilmark{10},
 E.~Morelli\altaffilmark{5}, E.~Moretti\altaffilmark{6},
A.~Morselli\altaffilmark{11}, L.~Pacciani\altaffilmark{1},
A.~Pellizzoni\altaffilmark{20}, F.~Perotti\altaffilmark{3},
G.~Piano\altaffilmark{1,2,11}, P.~Picozza\altaffilmark{2,11},
M.~Pilia\altaffilmark{21},  G.~Porrovecchio\altaffilmark{1},
G.~Pucella\altaffilmark{1}, M.~Prest\altaffilmark{12},
M.~Rapisarda\altaffilmark{13}, A.~Rappoldi\altaffilmark{7},
A.~Rubini\altaffilmark{1}, P.~Soffitta\altaffilmark{1},
M.~Trifoglio\altaffilmark{5}, A.~Trois\altaffilmark{1},
E.~Vallazza\altaffilmark{6}, V.~Vittorini\altaffilmark{1,2},
A.~Zambra\altaffilmark{3}, D.~Zanello\altaffilmark{9},
 P.~Santolamazza\altaffilmark{14},
P.~Giommi\altaffilmark{14}, S.~Colafrancesco\altaffilmark{14},
L.A.~Antonelli\altaffilmark{19},  L.~Salotti\altaffilmark{15}}

\altaffiltext{1} {INAF/IASF-Roma, I-00133 Roma, Italy}
\altaffiltext{2} {Dip. di Fisica, Univ. Tor Vergata, I-00133 Roma,
Italy} \altaffiltext{3} {INAF/IASF-Milano, I-20133 Milano, Italy}
\altaffiltext{4} {CIFS-Torino, I-10133 Torino, Italy}
\altaffiltext{5} {INAF/IASF-Bologna, I-40129 Bologna, Italy}
\altaffiltext{6} {Dip. Fisica and INFN Trieste, I-34127 Trieste,
Italy} \altaffiltext{7} {INFN-Pavia, I-27100 Pavia, Italy}
\altaffiltext{8} {ENEA-Bologna, I-40129 Bologna, Italy}
\altaffiltext{9} {INFN-Roma La Sapienza, I-00185 Roma, Italy}
\altaffiltext{10} {CNR-IMIP, Roma, Italy} \altaffiltext{11} {INFN
Roma Tor Vergata, I-00133 Roma, Italy} \altaffiltext{12} {Dip. di
Fisica, Univ. Dell'Insubria, I-22100 Como, Italy}
\altaffiltext{13} {ENEA Frascati,  I-00044 Frascati (Roma), Italy}
\altaffiltext{14} {ASI Science Data Center, I-00044
Frascati(Roma), Italy} \altaffiltext{15} {Agenzia Spaziale
Italiana, I-00198 Roma, Italy} \altaffiltext{16} {Osservatorio
Astronomico di Trieste, Trieste, Italy} \altaffiltext{17} {CRESST
and Universities Space Research Association, NASA/Goddard Space
Flight Center, Code 662, Greenbelt, MD 20771} \altaffiltext{18}
{Dip. Fisica, Universit\'a di Torino, Turin, Italy}
\altaffiltext{19} {INAF-Osservatorio Astron. di Roma, Monte Porzio
Catone, Italy} \altaffiltext{20} {INAF-Osservatorio Astronomico di
Cagliari, localita' Poggio dei Pini, strada 54, I-09012 Capoterra,
Italy} \altaffiltext{21} {Dipartimento di Fisica, Universit\'a
dell'Insubria, Via Valleggio 11, I-22100 Como, Italy}

%





\begin{abstract}

We present the results of extensive observations by the gamma-ray
AGILE satellite of the Galactic region hosting the Carina nebula
and the remarkable colliding wind binary  Eta Carinae (\etacarp)
during the period 2007 July -- 2009 January. We detect a gamma-ray
source (\aglp) consistent with the
position of \etacarp. If \agl is associated with the Car system
our data provide the long sought first detection above 100 MeV of
a colliding wind binary. The average gamma-ray flux above 100 MeV
and integrated over the
{\mtav pre-periastron} period 2007 July -- 2008 October is
$F_{\gamma} = (37 \pm 5) \times 10^{-8} \rm \, ph \, cm^{-2} \,
s^{-1} $ corresponding to an average gamma-ray luminosity of
$L_{\gamma} = 3.4 \times 10^{34} \rm \, erg \, s^{-1}$ for
a distance of 2.3 kpc. We also report a
2-day gamma-ray flaring episode of \agl on 2008 Oct.~11-13 {\mtavv
possibly}
related to {\mtavv a } transient acceleration and radiation
episode of {\mtavv the}  strongly variable {\mt shock in the
system}.


\end{abstract}





\section{Introduction}

$\eta$~Car is a
{\mtavv very} massive ($\sim100$ solar masses) star known for its
strong mass outflow eruptions, and is one of the most interesting
objects of our Galaxy (e.g., Davidson \& Humphrey 1997).
Radial velocity
variations of spectroscopic lines accumulated over the years
provide a strong evidence that \etacar is a  binary system (e.g.,
Damineli et al. 2008a,b) where the primary star is a luminous blue
variable (LBV) star orbiting in a very eccentric binary ($e \sim
0.9$) with a companion star believed to be an O star of $\sim 30$
solar masses.
The orbital period is 5.53 years ($\sim$2023
days) (e.g., Damineli et al., 2008a): the system has been
monitored in the radio, mm, IR, optical and X-ray
bands for at least three cycles.
Both stars emit dense and high-velocity gaseous winds, and the
binary system is ideal to study the interaction of colliding winds
and to test theories of particle acceleration and radiation under
extreme conditions. The mass outflow rates and wind speeds  of the
two stars inferred from the wealth of all available data are $
\dot{M_1} \simeq 2 \times 10^{-4} \, {\rm M_{\odot} \, yr^{-1}},
\dot{M_2} \simeq 2 \times 10^{-5} \, {\rm M_{\odot} \, yr^{-1}},
v_1 \simeq 600 \, {\rm km \, s^{-1}}, v_2 \simeq 3000 \, {\rm km
\, s^{-1}}$ \citep{pittard-2}.
%
{\mtav Eta Car is then interesting among other colliding wind
binaries since the observable X-ray emission in the 2-10 keV band
is almost entirely produced by the shocked fast wind of the
secondary star, with little if any contribution from the slow
shocked wind of Eta Car itself.}
 The  system is known for
its variability and occasional erratic eruptions detected in the
IR and optical bands, as well as for its distinct asymmetric
pattern of {\mtavv optical} line and X-ray emission during its
orbital period \citep{corcoran}.

\etacar has been  repeatedly observed in the energy ranges 1-10
keV and 20-100 keV by different observatories. It is certainly the
only source showing a non-thermal X-ray spectrum
 within a region
 centered on \etacar with a 1 degree diameter
 {\mt (the anomalous X-ray pulsar AXP~1E~1048.1-5937 is about 0.6 degrees away)}.
 Whereas the 1-10 keV spectrum is dominated by a quasi-thermal
and variable component \citep{corcoran,corcoran-2,viotti}, the
hard X-ray observations show non-thermal emission that appears to
vary along the orbit \citep{viotti-2,leyder}.
The \etacar source was
detected {\mtt with high significance}
 by BSAX-PDS and INTEGRAL-ISGRI
{\mt
{\mtavv far from } periastron}.
INTEGRAL is capable of resolving field sources with a few
arcminute resolution {\mtavv in the hard X-ray range. Although
INTEGRAL observed the system at different phase periods
(0.99-0.01, 0.16--0.19, 0.35-0.37),} \etacar was detected  {\mtav
only} during {\mtavv the phase interval} 0.16--0.19 with an
average 22-100 keV X-ray flux of $F = 1.1 \times 10^{-11} \rm \,
erg \, cm^{-2} \, s^{-1}$ \citep{leyder}.

The Carina region has been observed at gamma-ray energies above a
few MeV by the OSSE, COMPTEL and EGRET instruments on board of the
Compton Gamma-Ray Observatory (CGRO). An EGRET gamma-ray source
(3EG~J1048-5840)  is catalogued at about 1 degree distance from
\etacar. However, no gamma-ray emission above 100 MeV has been
reported by CGRO from the \etacar region\footnote{\mtav EGRET
observed the \etacar field several times (see, e.g., Table~2 of
Kaspi et al. 2000). Observations close to periastron were   VP32
(1992 Jun 25 - Jul 02, phase 0.014) and VP630 (1997 Sep 22 - Oct
06, phase 0.963) \cite{hartman,kaspi}.}.

The gamma-ray astrophysics mission AGILE \citep{tavani-1} observed
several times the Carina region in the Galactic plane during its
early operational phases and {\mtavv Cycle-1} observations. Here
we report the main results of the gamma-ray observations of the
\etacar region carried out by the AGILE satellite during the
period 2007 July - 2009 January, simultaneously in the energy
bands 30 MeV - 30 GeV and 18-60 keV. A {\mtavv high-confidence}
gamma-ray source (\aglp) was detected positionally consistent with
\etacar by integrating all data, as well as by considering
specific observation periods.


\section{AGILE 2007-2008 Observations}


 The AGILE mission has been operating since 2007 April
 \citep{tavani-1}.
The AGILE scientific instrument is very compact and is
characterized by two co-aligned imaging detectors operating in the
energy ranges 30 MeV - 30 GeV (GRID, Barbiellini et al. 2002,
Prest et al., 2003) and 18-60 keV (Super-AGILE, Feroci et al.
2007), as well as by an anticoincidence system \citep{perotti} and
a calorimeter \citep{labanti}. AGILE's performance is
characterized by large fields of view (2.5 and 1 sr for the
gamma-ray and hard X-ray bands, respectively), optimal angular
resolution and good sensitivity (see Tavani et al. 2008 for
details about the mission and main instrument performance). Flux
sensitivity for a typical 1-week observing period can reach the
level of several tens of $ 10^{-8} \rm \, ph \, cm^{-2} \, s^{-1}
$ above 100 MeV, and 10-20 mCrab in the 18-60 keV range depending
on off-axis angles and {\mtav pointing directions.}

 \begin{table*}
\begin{center}
\small \caption{AGILE observations of the \etacar region}
\begin{tabular}{|c|c|c|c|c|c|c|c|}
  \hline
  AGILE orbits & Date interval & MJD & N. days & \etacar  &  $\sqrt{TS}$ (**) & Counts
  & Average flux(***) \\
  &  & &  &  phase (*) &   &
  &  \\[0.5ex]
  \hline
1146-1299 & 2007.07.13-2007.07.24 & 54294.5-54305.5 & 11 & 0.732& 4.1 & 105 $\pm$ 29 & 67 $\pm$ 18\\
1429-1567 & 2007.08.02-2007.08.12 & 54314.5-54324.5 & 10 & 0.741 & 3.3 & 80 $\pm$ 27 & 58 $\pm$ 20\\
1584-1708 & 2007.08.13-2007.08.22 & 54325.5-54334.5 & 9 & 0.747& 1.1 & $<$ 73  &  $< 67 $ \\ 
3673-4009 & 2008.01.08-2008.02.01 & 54473.5-54497.5 & 24 & 0.824& 4.1 & 166 $\pm$ 45 & 55$\pm$ 14\\ 
4194-4418 & 2008.02.14-2008.03.01 & 54510.5-54526.5 & 16 & 0.840& 3.1 & 96 $\pm$ 33 & 43 $\pm$ 15 \\
6129-6480 & 2008.06.30-2008.07.25 & 54647.5-54672.5 & 25& 0.910& 5.6 & 214 $\pm$ 42 & 61 $\pm$ 12 \\
6778-7001 & 2008.08.15-2008.08.31 & 54693.5-54709.5 & 16& 0.930& 1.3 & $<$ 105 & $< 46$\\ 
7569-7664 & 2008.10.10-2008.10.17 & 54749.5-54756.5 & 7 & 0.956& 4.0 & 80 $\pm$ 23 & 99 $\pm$ 28 (****) \\
8899-8995 & 2009.01.12-2009.01.19 & 54843.5-54850.5 & 7 & 0.002 & 2.2 & $48 \pm 22$  & $< 94$  \\
\hline
\end{tabular}
\end{center}
 \label{tab-1}

 \vspace*{1.cm}
 (*) Average orbital phase of \etacar calculated at the center of
 the time interval.\\
\noindent {\mtt (**)  Square root of the the Maximum Likelihood
Test Statistic (TS) representing the statistical significance of
the detection.}\\
\noindent (***)  Gamma-ray flux of \agl above 100 MeV in units of
$10^{-8} \,\rm  ph \, cm^{-2}$ s$^{-1}$ obtained by taking into
account the nearby source \agll in the multisource
 likelihood analysis.
 { We also indicate 2-sigma upper limits in the same units}.\\
 (****) {\mtav During this period the source reached the
 gamma-ray flux above 100 MeV of $F = (270 \pm
65) \times 10^{-8} \rm \, ph \, cm^{-2} \, s^{-1} $.}
\end{table*}


The AGILE satellite  repeatedly pointed at the Carina region
{\mtav for a total of $\sim130$ days } during the time period 2007
July - 2009 January. Table~1 summarizes the AGILE observations of
the field.
 Analysis of the
gamma-ray data was carried out with the $FT3ab_2$
calibrated filter, with a gamma-ray event selection that takes
into account standard SAA event cuts and 80 degree Earth albedo
filtering. We used the AGILE gamma-ray software release Build 17
and {\mt the standard} hard X-ray analysis software.

\begin{figure} 
\begin{center}

 \includegraphics [height=6.5cm]{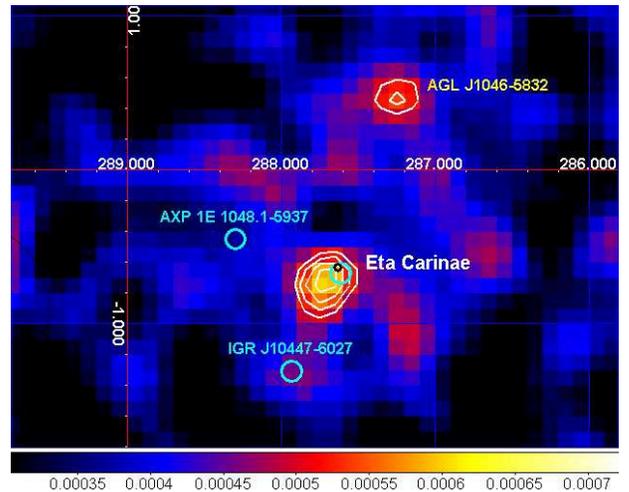}

\caption{\mt AGILE gamma-ray intensity map in Galactic coordinates
of the $\eta$ Car region above 100 MeV summing all data collected
from 2007 July to 2008 October. The central gamma-ray source that
can be associated with $\eta$ Car is \aglp; we also indicate the
prominent nearby gamma-ray source \agll which is associated with
the radio pulsar PSR~B1046-58 (Kaspi et al. 2006; Abdo et al.
2009). The color bar scale is in units of $\rm photons \, cm^{-2}
\, s^{-1} \, pixel^{-1}$. Pixel size
is 0.1 degrees, and we used a 3-bin Gaussian smoothing. White
contour levels of the AGILE sources start from 0.0005 {\mtavv and
increase} in steps of 0.000028. The optical position of \etacar is
marked by a small black circle. The INTEGRAL sources (Leyder et
al. 2008) are marked with cyan circles. } \label{etacar-fig-3}
\end{center}
\end{figure}

 \begin{figure}[h!]
\begin{center}
 \includegraphics [width= 8.5cm, height=6.2cm]{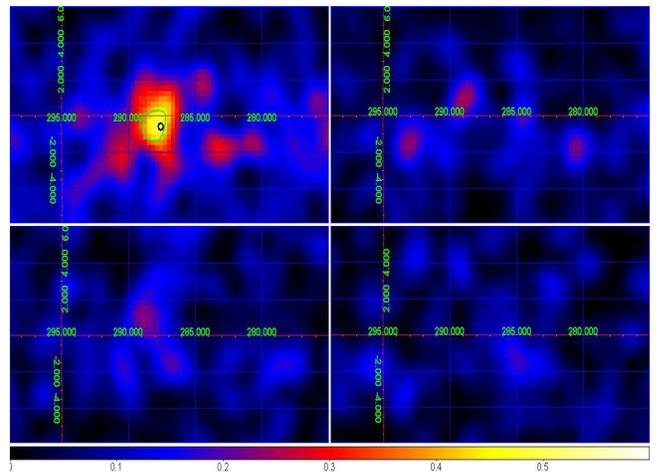}
\caption {Time sequence of AGILE-GRID gamma-ray counts maps above
100 MeV centered on the \etacar region during the period 2008
10-17 October.  The time sequence starts at the upper right corner
and it is counterclockwise. Each map corresponds approximately to
a 2-day integration starting on 2008 Oct. 10. {\mt The color bar
scale is in units of $\rm counts \, cm^{-2} \, s^{-1} \,
pixel^{-1}$. Pixel size 
is 0.3 degrees for a 3-bin Gaussian
smoothing.} A  strong gamma-ray {\mt
flare} 
shows up in the second map for the period  2008
Oct. 11 (02:57 UT) - 2008 Oct. 13 (04:16 UT). The green contour
marks the 95\% contour level error box. The position of the
gamma-ray 
{\mt flaring source} is consistent with the \agl position and with
\etacar (marked by a black small circle).  } \label{etacar-fig-4}
\end{center}
\end{figure}

Fig.~\ref{etacar-fig-3} shows the integrated sky map of the
\etacar region at energies above 100 MeV
for the
period 2007 July - 2008 October. A gamma-ray source is
detected with high confidence
(7.8 sigma) at the position
$(l,b) =  287.6, - 0.7 \pm 0.3 \rm (stat.) \pm 0.1 (syst.)$. The
average gamma-ray flux above 100 MeV and integrated over the whole
period 2007 July - 2008 October is $F_{\gamma} = (37 \pm 5) \times
10^{-8} \rm \, ph \, cm^{-2} \, s^{-1} $. We call this
source\footnote{\mtavv The same source is also listed in the Fermi
Bright Source Catalogues  as 0FGL~J1045.6-5937 (Abdo et al.
2009).} \agl
{\mtavv following the source designation of } the first AGILE
catalog of high-confidence gamma-ray sources (Pittori et al.
2009). We included in all our multisource analysis the nearby
gamma-ray source\footnote{\mtav This source, which
{\mtavv did not reach the stringent} significance threshold of the
First AGILE Catalog \citep{pittori}, {\mtavv is included in the
Fermi Bright Source List as 0FGL~J1047.6-5834 and is identified
with PSR J1048-5832 (Abdo et al. 2009)}.}
\agll that the AGILE-GRID detects with an average and
constant gamma-ray flux above 100 MeV of
 $f_{\gamma} =
(27 \pm 4) \times 10^{-8} \rm \, ph \, cm^{-2} \, s^{-1} $.
\etacar is well within the 95\% confidence radius gamma-ray error
box of \aglp; the other nearby hard X-ray sources in the field
(the anomalous X-ray pulsar AXP~1E~1048.1-5937, and
IGR~J10447-6027) are excluded.  The Super-AGILE hard X-ray imager
did not detect a source coincident with \agl for both short and
long integrations of Table~1. {\mt Depending on the source
position in the FOV, the typical {\mt 3-sigma} Super-AGILE upper
limit is 10-20 mCrab, i.e., consistent with the INTEGRAL detection
{\mtavv and upper limits} of \etacarp.}


 We searched for short timescale variability of the
gamma-ray and hard X-ray flux from \agl throughout the  whole
AGILE observing periods of the Carina region. A  2-day {\mt
gamma-ray} {\mt flare} from the direction of \etacar was detected
during the observation
 of 2008 October 10-17. The
emission reached its peak gamma-ray emission during the period
2008 Oct. 11 (02:57 UT) - 2008 Oct. 13 (04:16 UT).
Our analysis gives a 5.2 sigma detection of a source at the
position $(l,b) = 288.0, -0.4 \pm 0.6$ {\mtavv fully} consistent
with the \agl position
{\mtavv but }
with a gamma-ray flux above 100 MeV of $F = (270 \pm
65) \times 10^{-8} \rm \, ph \, cm^{-2} \, s^{-1} $.
 Fig.~\ref{etacar-fig-4} shows the time sequence of 2-day integration
gamma-ray maps of the region during the period 2008 10-17 October.

\begin{figure} 
\begin{center}
\includegraphics [width= 9.cm, height=6.5cm]{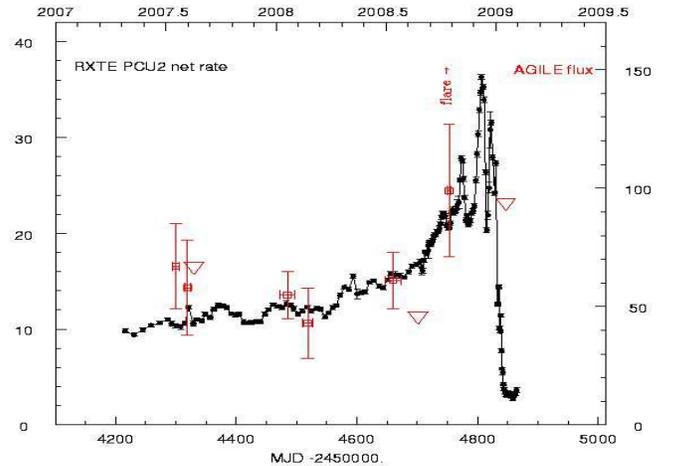}
\caption {AGILE gamma-ray lightcurve of \agl showing the fluxes
above 100 MeV (right axis scale in units of $10^{-8} \rm ph \,
cm^{-2} \, s^{-1}$) averaged over the observing periods of Table~1
(red crosses) and superimposed with the RXTE PCU2 net rate X-ray
light curve of \etacar (black symbols, left axis) obtained during
the dedicated campaign observing the last cycle and periastron
passage. {\mt Triangles indicate 2-sigma upper limits.} {\mtavv We
also mark the occurrence of the 2008 October 11-13 flare  when the
source reached a gamma ray flux above 100 MeV of $F = (270 \pm 65)
\times 10^{-8} \rm \, ph \, cm^{-2} \, s^{-1}$.}}
\label{etacar-fig-5}
\end{center}
\vspace*{0.5cm}
\end{figure}

Fig.~\ref{etacar-fig-5} shows the AGILE gamma-ray data of Table~1
superimposed with the simultaneous RXTE  (PCU2 net rate)
lightcurve in the energy band 2-15 keV during the period 2007
February - {\rv 2009 January}
(the typical and relatively abrupt
decrease of the X-ray emission near periastron is clearly
visible).

\begin{figure}[h!]
\begin{center}
 \includegraphics[width= 9.5cm, height=6.5cm]{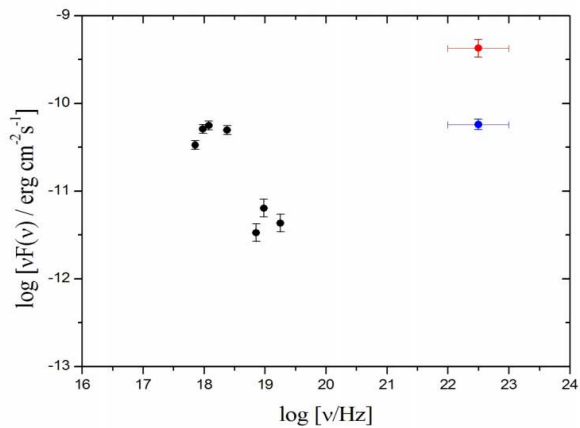}
 \caption {Combined spectral power {\mt flux} of \etacar  as reported by the
 SAX-MECS {\mtt in the energy range 1-10 keV (phase 0.46)} \citep{viotti-2}, {\mtt by }
 INTEGRAL {\mtt in the energy range 22-100 keV (phase 0.16-0.19) } (Leyder et al.
 2008) plotted together with the two {\mt broad-band} {\mtt gamma-ray} spectral states of
 \agl measured by AGILE during the period 2007 July -- 2008 October {\mtt (\etacar
phase 0.73-0.95)}. The lower blue point marks the average {\mtt
gamma-ray } spectral flux, and the upper red point indicates the
spectral state during the
gamma-ray {\mt flare } of 2008 October 11-13. }
\label{etacar-fig-spectrum}
\end{center}
\end{figure}

Fig.~\ref{etacar-fig-spectrum} shows two representative different
{\mt broad-band} spectral states of \agl  {\mt obtained} during
the period 2007 July -- 2008 October together with the historical
X-ray and hard X-ray data reported from \etacarp. We mark in blue
the average spectrum obtained by integrating all data {\mtt
outside periastron} , and with a red cross the flux corresponding
to the flaring state of 2008 October 11-13. We also report in the
same plot the {\mtt (non simultaneous)} BSAX-MECS and the
INTEGRAL-ISGRI spectral states of \etacar reported in the
literature \citep{viotti-2,leyder}. It is interesting to note that
if \agl is associated with \etacar, the average AGILE spectrum
together with the INTEGRAL historical spectrum is in qualitative
agreement with expectations based on inverse Compton and/or pion
decay models of gamma-ray emission from colliding wind binaries
(e.g., Reimer, Pohl, Reimer 2006). For the 2008 11-13 October
flaring episode, the Super-AGILE 18-60 keV upper limit is 70 mCrab
in the energy band 18-60 keV.
{\mtt Obtaining simultaneous hard X-ray and gamma-ray data during
the flaring state of \agl is crucial to study the broad-band
variability of the source. However, due to unfavorable source
positioning of \agl in the Super-A field of view in mid-October,
2008, the hard X-ray upper limit is not very constraining. }

AGILE pointed at the Carina Region during the period 2009 12-19
January as a special repointing to
{\mtavv cover} the \etacar periastron passage (calculated to be
occurring on 2009 January 11). {\mt We cannot exclude, at this
stage, the existence of a weak gamma-ray source consistent with
\aglp. However,
we
can currently provide a 2-sigma upper limit to the emission above
100 MeV of  $ 94\times 10^{-8} \rm \, ph \, cm^{-2} \, s^{-1} $.}
A more detailed analysis of the AGILE and multifrequency data
during the \etacar
 periastron passage will be discussed elsewhere.


\section{Discussion }


\etacar is {\mtav located} in the Carina nebula that extends for
several degrees in a Galactic region characterized by dense
molecular clouds, young stars and star formation sites.
However, within the \agl error box \etacar itself is by far the
strongest and hardest source in the 2-10 keV and 22-100 keV
ranges. {\mtt Another source inside the AGILE error box and } 7
arcmin away from
\etacar is the
X--ray
binary HD 93162/WR 25 (WN6+O4, with an orbital period of 208 days,
Gamen et al. 2007). This system {\rv is known} to be a
colliding wind system. Pollock \& Corcoran (2006) found
significant variability, possibly periodic, of its X--ray flux.
However, WR 25 {\mt was}  not detected in the hard X-ray range by
INTEGRAL during any of its observations. Furthermore, no other
prominent hard X-ray source is known in the \agl error box except
for \etacar itself. The two nearby hard X-ray sources detected by
INTEGRAL \citep{leyder} are outside the 95\% confidence level
error box {\rv of AGILE. Multiple gamma-ray sources within the
\agl error box cannot be excluded but are unlikely.}

Based on the accumulated multifrequency evidence
and the nature of the source, we consider the association of \agl
and \etacar as very likely. We briefly elaborate below on the
theoretical implications of our results assuming that \agl is
indeed the gamma-ray counterpart of \etacarp.


Colliding wind binaries (CWBs) are ideal systems to test theories
of hydrodynamical shocks and particle acceleration under extreme
radiative conditions provided by the proximity of the two stars.
In particular, supersonic winds can form efficient shocks where
electrons and protons can be accelerated through first-order Fermi
(Eichler \& Usov 1993) or other acceleration mechanisms. Inverse
Compton {\rv (IC)} scattering of shock-accelerated particles in
the presence of the very intense IR-optical-UV background of the
nearby
{\mt very bright } stars provides a crucial ingredient in CWBs. In
addition to synchrotron and Bremsstrahlung electron emissions, the
IC emission can dominate the high energy spectrum at energies
larger than several tens of keV up to MeV-GeV energies.
Furthermore, if protons are efficiently accelerated, they can
interact with the dense stellar outflows and produce gamma-rays by
pion production and neutral pion decay (e.g., Eichler \& Usov
1993, Benaglia \& Romero 2003; Reimer et al. 2006). All these
ingredients are important for the \etacar system and detailed
hydrodynamical modelling of the mass outflow have been developed
\citep{parkin,parkin-2,okazaki}. A comprehensive and detailed
theoretical analysis of our data is beyond the scope of this
paper. We  outline here a few important points.

If \agl is the \etacar gamma-ray counterpart, our data show the
first
remarkable detection of a colliding wind system at hundreds of MeV
energies, confirming the efficient particle acceleration and the
highly non-thermal nature of the strong shock in a CWB. The
average gamma-ray flux of \agl
{\mtavv translates into } gamma-ray luminosity of $L_{\gamma} =
3.4 \times 10^{34} \rm \, erg \, s^{-1}$ for an \etacar distance
of
2.3 kpc, 
{\mtavv corresponding } to a fraction of a percent of the total
wind kinetic power. The 2008 Oct. 11-13 flare episode has  a
luminosity of $L_{\gamma} = 2. \times 10^{35} \rm \, erg \,
s^{-1}$.  The average {\mt broad-band} gamma-ray spectrum
determined by AGILE is in qualitative agreement with expectations
of CWB spectra as calculated for dominant IC and neutral pion
decay processes \citep{benaglia,reimer}.

{\mtavv  While the gamma-ray flux of \agl is roughly constant
during the time span covered by our observations, a significant
variability was detected on a few day time-scale in October 2008.
This episode
indicates that}
the gamma-ray emission can be {\mt associated with}
intermittent strong shock acceleration episodes and/or magnetic
field enhancements to be expected for a very variable and
inhomogeneous mass outflow from the stars of the \etacar system.
%
%
In particular, we note that the strong gamma-ray flaring episode
occurred a few months before periastron, when the efficiency of
transforming a mass outflow enhancement into
{\mtavv particle acceleration is expected to increase
because of} the closeness of the two stars. \etacar provides then
some crucial ingredients regarding the formation of high-energy
emission in CWBs: (1) strong variability of the mass outflows; (2)
a high-speed wind from the less massive companion; (3) a radiative
environment with a specific bath of soft photons from both stars
(IR, optical and UV fluxes) that can illuminate the shock region
and provide a time variable environment for enhanced IC emission
in the 100 MeV range and beyond. The theoretical implications are
far reaching. The \etacar system would provide the first CWB to
test the particle acceleration models for non-relativistic mass
outflows under a specific set of physical conditions. It is very
important to assess the efficiency of the particle acceleration
process in such a radiative environment. A gamma-ray flaring
episode lasting $\sim2$ days implies a fast acceleration timescale
and subsequent radiation and decay of the strong shock properties
leading to the efficient emission.
%
If the gamma-ray emission is associated with \etacar our
observations provide  important data to test shock acceleration
models. {\mtav Future gamma-ray observations and analysis will
further contribute to enlighten the emission mechanism and the
ultimate origin of \aglp.}

\end{document}